 \newlength\smallfigwidth
 \documentclass[preprint,twocolumn,floatfix,showpacs,amsmath,amssymb]{revtex4}

\usepackage{graphicx}
\usepackage[hypertex]{hyperref}
\begin{document}

\preprint{UFV}

\title{Scattering of charge carriers in graphene induced by topological defects}

\author{J.\ M.\ Fonseca}
\author{A.\ R.\ Pereira}
\email{apereira@ufv.br}
\author{W.\ A.\ Moura-Melo}
\affiliation{Departamento de F\'isica, Universidade Federal de
Vi\c cosa, Vi\c cosa, 36570-000, Minas Gerais, Brazil}

\date{}

\begin{abstract}
We study the scattering of graphene quasiparticles by topological defects, represented by holes, pentagons and heptagons. For holes, we found that at low concentration they give a negligible contribution to the resistivity. Whenever pentagons or heptagons are introduced we realize that a fermionic current is scattered by defects.
\end{abstract}

\maketitle

\begin{figure}
\includegraphics[angle=0.0,width=8cm]{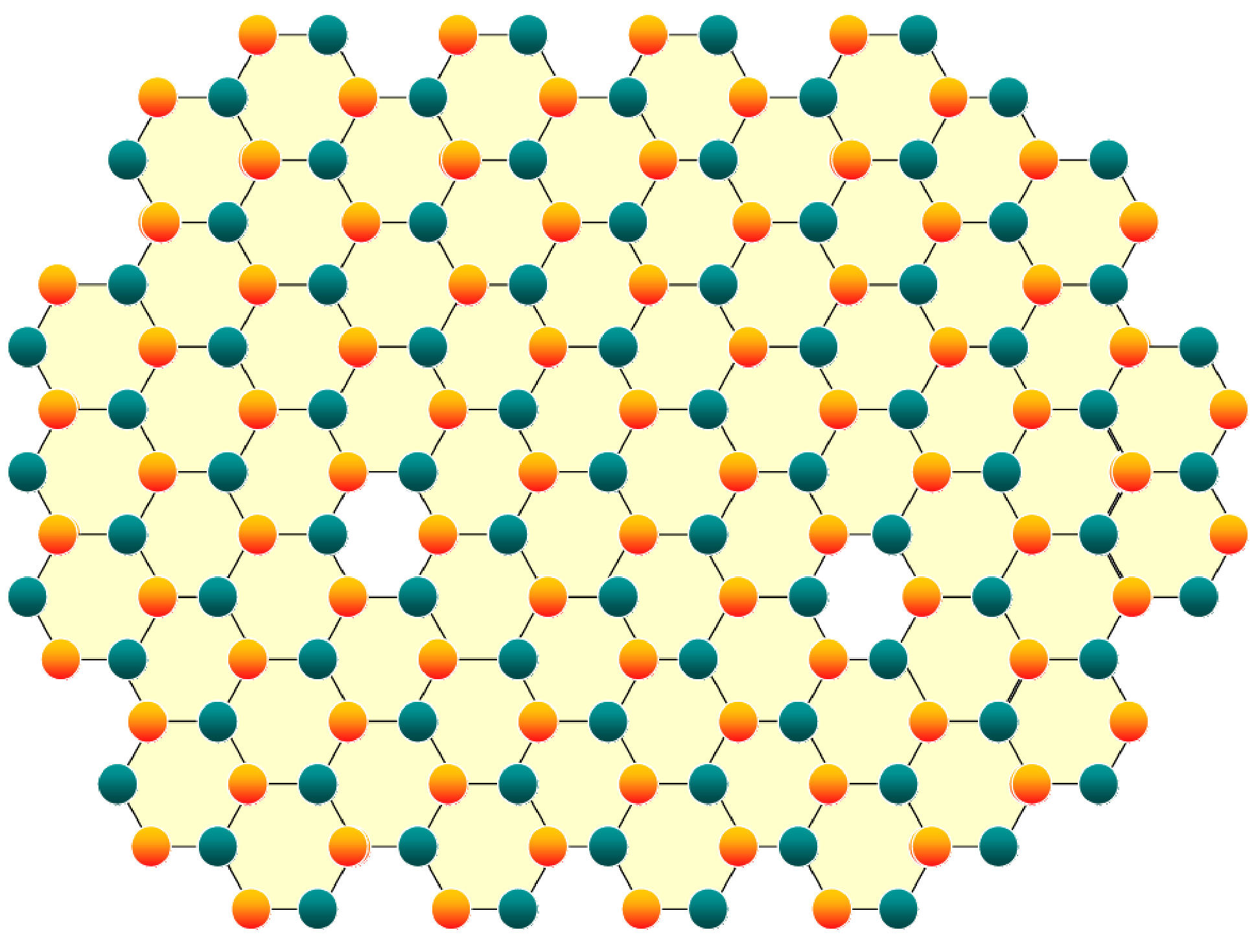}
\caption{ \label{figplano} (Color online) Graphene is a flat
monolayer of carbon atoms having a honeycomb lattice, consisting
of two interpenetrating triangular sublattices (green and red
circles). Experimental techniques provide high-quality graphene
crystallites up to $100\mu m $ in size, which is sufficient for
most research purpose, including the one considered in this
letter. The charge carries can travel thousands of interatomic
distances without scattering.}
\end{figure}

\indent  The study of lower dimensional gravity is generally
considered to be only an ``academic" exercise, providing an useful
arena for developing new ideas and insights, which are applicable
to classical and even quantum theory of gravitation in $3+1$
dimensions [\onlinecite{Brown88}]. In this letter we would like to
suggest that the Einstein theory of gravitation in $2+1$
dimensions can be seen and even tested experimentally in possible
realizable condensed matter materials, such as graphene. To
understand our point of view, it would be instructive to review
some properties of this material. As well known, graphene is a
flat monolayer of carbon atoms tightly packed into a
two-dimensional ($2D$) honeycomb lattice (see Fig.\
\ref{figplano}) , or it can be viewed as an individual atomic
plane pulled out of bulk graphite
[\onlinecite{Novoselov04,Novoselov05}]. It is the first example of
truly atomic $2D$ crystalline matter and is the basic building
block for graphitic materials such as fullerenes (graphene balled
into a sphere) or carbon nanotubes (graphene rolled-up in
cylinders) [\onlinecite{Geim07,Castro-Neto06,Geim08}]. From the
point of view of its electronic properties, graphene is a zero-gap
semiconductor, in which the low energy spectrum is formally
described by the Dirac-like equation for a massless particle
[\onlinecite{Novoselov05,Katsnelson06,Wallace47}]
\begin{eqnarray}\label{Dirac}
i\hbar\frac{\partial}{\partial t}|\Psi\rangle =
v_{F}\vec{\sigma}\cdot \vec{p} |\Psi\rangle,
\end{eqnarray}\\
where $v_{F}$ is the Fermi velocity, which plays the role of the
speed of light ($v_{F}\approx c/300$), $\vec{\sigma}=(\sigma_{x},
\sigma_{y})$ are the $2D$ Pauli matrices, $\vec{p}=-i\hbar
\vec{\nabla}$ is the linear momentum operator and $|\Psi\rangle$
is a two-component spinor. Therefore, the quasiparticles (called
massless Dirac fermions) can be seen as electrons that have lost
their mass or as neutrinos that acquired the electron charge. In
the honeycomb lattice of graphene, the two-component spinor
$|\Psi\rangle$ is referred to as pseudospin since it is an index
to indicate two interpenetrating triangular sublattices $A$ e $B$,
which is similar to spin index (up and down) in quantum
electrodynamics (QED). It is common to regard the sublattice
degree of freedom as a pseudospin, with the $A$ sublattice being
the ``up" ($|+ \rangle$) pseudospin state and $B$ sublattice being
the ``down" ($|- \rangle$) pseudospin state, i.e.,
\begin{eqnarray}\label{ABphase}
|+ \rangle=\left(
  \begin{array}{cc}
    1 \\
    0 \\
  \end{array}
\right) ;  |- \rangle=\left(
  \begin{array}{cc}
    0 \\
    1 \\
  \end{array}
\right).
\end{eqnarray} \\
Since $v_{F}\ll c$, it is a slow relativity system or a strong
coupling version of QED since the graphene's dimensionless
coupling constant ($e^{2}/\hbar v_{F}$) is on the order of unit,
much larger than that of QED (the fine structure constant
$e^{2}/\hbar c\approx 1/137$). All these properties make the
graphene a very interesting system, which provides a way to probe
QED phenomena by measuring its electronic properties. Several
proposals for testing some predicted but not yet observed
phenomena in QED, including topics related to the Klein paradox
[\onlinecite{Katsnelson06, Katsnelson07}], issues of vacuum
polarization [\onlinecite{Shytov07}] and atomic collapse
[\onlinecite{Shytov+07}], etc, are under investigation in
graphene. It becomes more exciting when one remembers that these
probes are made in a low energy condensed matter system. All these
potential connections between high energy and condensed matter
physics inspire us to consider another question: could these
strictly $2D$ materials actuate as a bridge between condensed
matter physics and lower dimensional gravitation [more
specifically, three-dimensional ($2+1$) gravitation]? It is not an
empty problem because freely suspended graphene membrane is
partially crumpled [\onlinecite{Meyer07}] and hence, it would be
also interesting to study Dirac fermions in curved space. Here we
will consider one of the simplest curved space, which is
associated with the Schwarzschild solution in $2+1$ dimensions: a
space locally flat (with the Minkowski metric) but globally
nontrivial (as described below). In this case, graphene would
provide an appealing way to experimentally explore general
relativity (GR) in two spatial dimensions.

\begin{figure}
\includegraphics[angle=0.0,width=8cm]{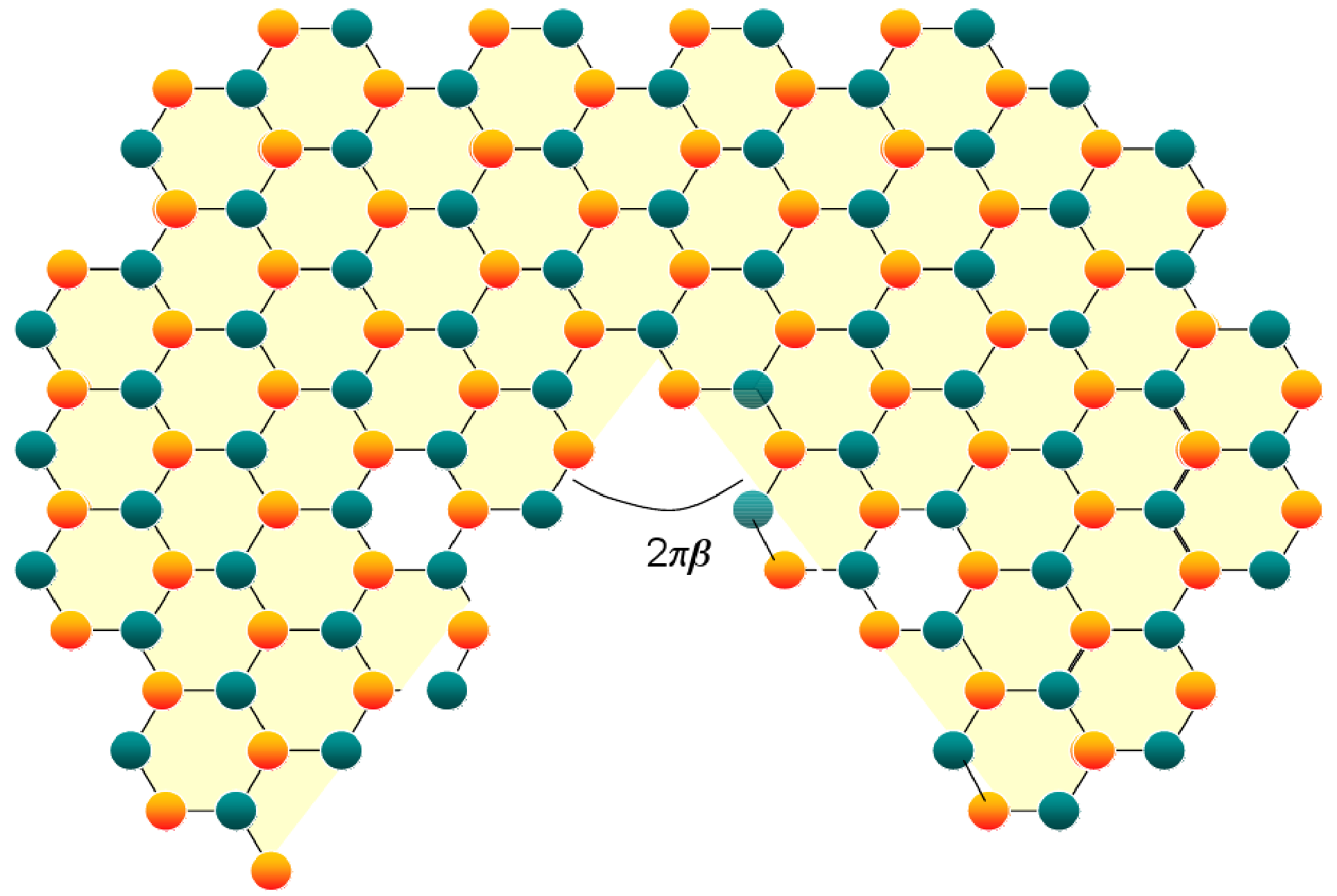}
\caption{ \label{figremove} (Color online) Detaching a fraction of
a graphene sheet to generate a deficit angle $2\pi \beta$.}
\end{figure}

\begin{figure}
\includegraphics[angle=0.0,width=8cm]{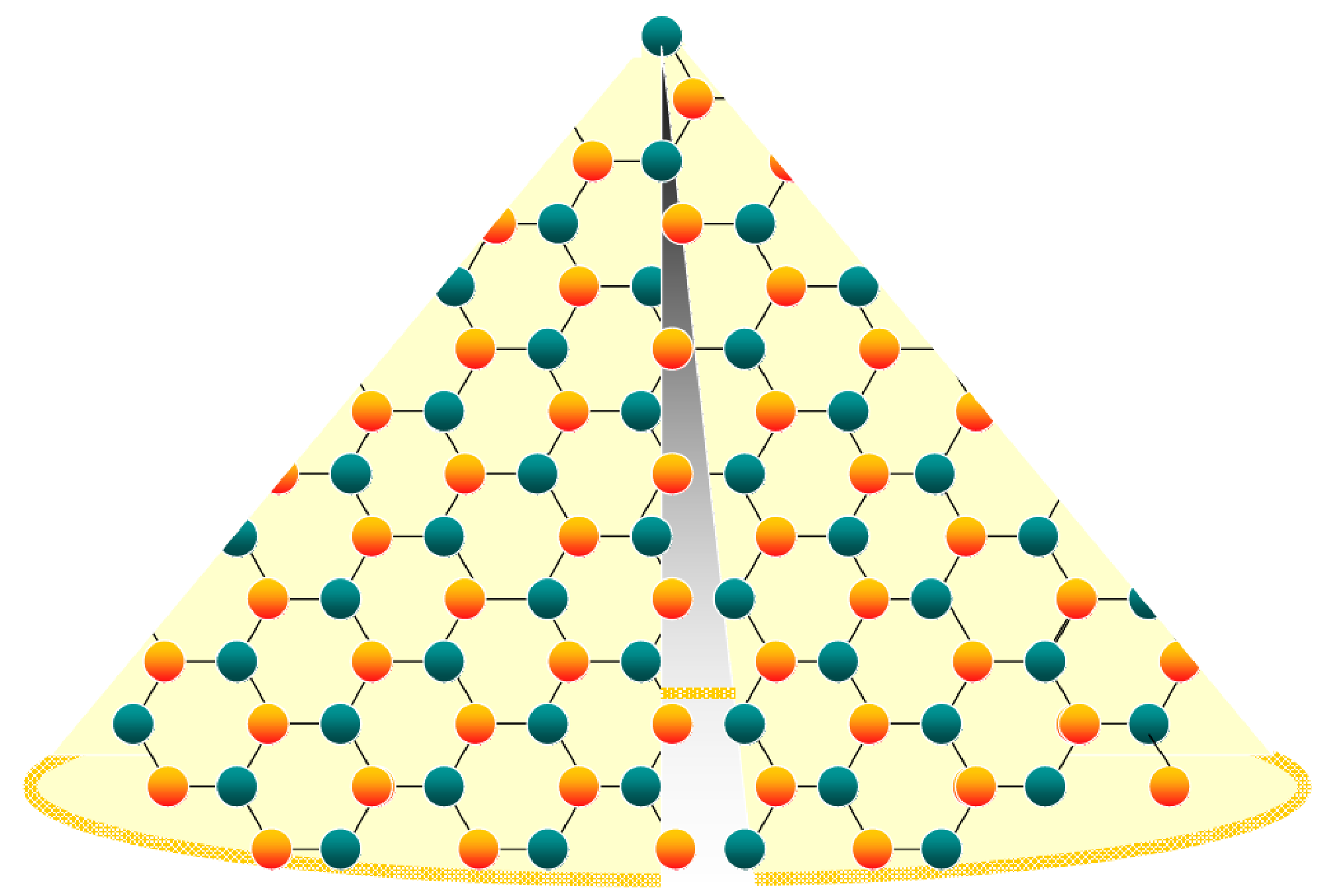}
\caption{ \label{figcone}  (Color online) If a wedge is removed
from the graphene (see Fig. \ \ref{figremove}) and edges
identified, a cone results. The motion of the charge carriers in
an ideal conical graphene is equivalent to that of a massless
Dirac particle in a gravitational field of a static particle of
mass $M$ in a $2+1$ dimensional space-time.}
\end{figure}

To go further with our connection between graphene and gravitation
we now summarize some aspects of general relativity in three
space-time dimensions. It differs fundamentally from that of four
dimensions. Indeed, it exhibits some unusual features, which can
be deduced from the properties of the Einstein field equations and
the Riemann curvature tensor $R^{\mu}_{\upsilon \varepsilon
\kappa}$ [\onlinecite{Brown88,Staruszkiewikz63}]. In regions which
are free of matter (or wherever the momentum-energy tensor $T^{\mu
\nu}$ vanishes), the space-time is locally flat when the
cosmological constant is zero (the Einstein tensor $G^{\mu\nu}$
vanishes). However, this does not mean that a pointlike source has
no gravitational effects: a light beam passing by a massive,
pointlike source will be deflected
[\onlinecite{Staruszkiewikz63,Deser84,Giddings84,Vilenkin81}] and
parallel transport in a closed circuit around such a source will
in general give nontrivial results [\onlinecite{Burges85}].
Indeed, while the local curvature vanishes outside the sources,
there are still nontrivial global effects. For instance, for the
special case of a pointlike mass $M$ sitting at rest in the
origin, the line element is given by
\begin{eqnarray}\label{metric}
ds^{2}= dt^{2}-d\rho^{2}-\rho^{2}\alpha^{2}d\theta^{2},
\end{eqnarray}\\
with $0\leq \theta < 2\pi $ and $\alpha=1-4GM$ ($G$ is the
Newton's constant, which has dimensions of $length$ rather than
$length^{2}$ if ones takes the natural units $h=c=1$). Note that
although the situation looks trivial, the coordinate $\theta$
ranges from $0$ to $2\pi \alpha$, indicating a deficit angle in
space. Then, the space part of the metric is that of a plane with
a wedge removed and edges identified; the unique $2D$ spatial
geometry satisfying this description is the cone
[\onlinecite{Staruszkiewikz63}]. This is the point we would like
to explore after the existence of $2D$ crystals whose excitations
are massless Dirac fermions. Our proposal for association of
graphene and gravitation is outlined in two parts: firstly,
exploring the $2D$ character and flexibility of this material, the
idea is to make a system in which one or more sectors are excised
from a graphene (see Fig.\ \ref{figremove}) and the remainder is
joined seamlessly (Fig.\ \ref{figcone}). In fact, the missed link
of each carbon atom resting at the two edges of the remaining
graphene sheet can be, in principle, covalently bounded. The
nucleation and growth of curved carbon structures are still no
well understood. It seems that the occurrence of pentagons, which
yield $60^{\circ}$ disclination defects in hexagonal graphitic
network, is a key element in the puzzle. Particularly, considering
the symmetry of a graphite sheet and the Euler's theorem, it can
be shown that only five types of cones (incorporating one to five
pentagons) can be made of a continuous sheet of graphite.
Graphitic cones were already produced and all the types were
synthesized [\onlinecite{Ge94,Krishnan97}]. For our purposes, it
would be ideal that the incorporation of pentagons (if necessary)
should be minimum (one to five pentagons may be tolerable) in the
construction of a cone with a graphene sheet. Indeed, only a
honeycomb lattice produces the peculiar energy-momentum
relationship $E=v_{F}p$ and, therefore, pentagons may cause
additional effects in the charge carries beyond the one analyzed
here. Secondly, with such a material one could study the
properties of the massless Dirac fermions in a space which is
locally described by the Minkowski metric but, however, possess
nontrivial global effects that should affect the physical
properties of these quasiparticles, bringing consequences to the
electronic properties of the sample as whole. As a first
discussion about this, we consider the changes that a closed-path
parallel transport around the cone tip causes in the Dirac spinors
$|\Psi\rangle$. Basically, it results in the Aharonov-Bohm effect
but it does not occur in the presence of an electromagnetic field.
Its origin is a defect in the material (the cone tip) which is
very similar to the gravitational field generated by a static
particle of mass $M$ if the space were two-dimensional instead
three-dimensional. Therefore, it will be called gravitational
Aharonov-Bohm effect in graphene. Such phenomenon could also be
used to probe GR in two spatial dimensions.

In the case of a conical graphene, it will be useful to substitute
the gravitational term $4GM$ by the symbol $\beta$, which gives a
deficit angle equal to $2\pi \beta$. It implies that the cone
surface makes an angle $\delta=\arcsin(\beta)$ with its symmetry
axis (the $z$-axis, which passes by the cone tip). The possible
five graphitic cones mentioned earlier are then given by
$2\delta=19.2^{\circ},38.9^{\circ},60^{\circ},84.6^{\circ},112.9^{\circ}$.
Our aim is, therefore, to see the influences that such a special
graphene structure could induce on quasiparticle wavefunctions
(spinors); surely, these influences may create new perspectives in
the electronic transport properties, which are determined by the
quasiparticles constrained to move in the conical surface. Hence,
the first thing to do in this direction, is to examine the effects
of closed-path parallel transport around the cone tip in the Dirac
quasiparticle wavefunctions. It can be obtained by integrating the
equations of parallel transport directly [\onlinecite{Burges85}]
or by using the path-ordered products of the relevant affine
connection [\onlinecite{Bezerra87}]. Thus, the two-component
spinors representing the pseudospin in graphene is affected by
parallel transport as follows
\begin{eqnarray}\label{SpinorClosed}
|\Psi (2\pi)\rangle = [\cos (\pi \sin \delta)-i \sigma_{z}
\sin(\pi \sin \delta)]|\Psi(0)\rangle,
\end{eqnarray}\\
where $\sigma_{z}$ is the Pauli matrix. Therefore, for a cone made
with a single graphene sheet, the quasiparticle wavefunctions
would experience an Aharonov-Bohm phase given by
\begin{eqnarray}\label{ABphase}
|\Psi (2\pi)\rangle=\left(
  \begin{array}{cc}
    \exp(-i\phi) & 0 \\
    0 & \exp(i\phi)\\
  \end{array}
\right) |\Psi (0)\rangle,
\end{eqnarray} \\
with $\phi=\pi \sin\delta$. Note that the spinors are very
sensitive for detecting point defects in condensed matter (or
equivalently, pointlike masses in gravitation). For instance, if
$|\Psi (0)\rangle = p|+ \rangle + q|- \rangle$ ($p$ and $q$ are
constants), then, $|\Psi (2\pi)\rangle = p \exp(-i\phi)|+ \rangle
+ q \exp(i\phi)|- \rangle$ and, therefore, after the complete
contour, the quasiparticles of $A$ and $B$ sublattices are
mutually out of phase by $2\pi \sin\delta = 2\pi\beta$, which is
exactly the deficit angle. Hence, the angular defect may give rise
to a mismatch of the transport properties of electrons and holes
in the $A$ and $B$ sublattices. Note that the phase difference in
the spinor components has the required property of vanishing when
the space becomes a plane, i.e., $\delta=0$ ($\alpha=1$). On the
other hand, if $\delta \neq 0$, the electrons of one sublattice
will be advanced in relation to the electrons of the other
sublattice. This effect may manifest itself in fluctuations or
concentration of charge carriers, which in turn alters several
physical properties. For instance, in a planar graphene, it is
known that the hopping of electrons between sublattices produces
an effective magnetic field that is proportional, in magnitude and
direction, to momentum measured from the Brillouin-zone corners.
This effective field, which acts on the pseudospin, may suffer
important changes in the conical graphene because a hopping of a
quasiparticle, which was previously, say, in the $A$ sublattice,
will make it to become out of phase with all quasiparticles
already occupying the $B$ sublattice. Thus, if it is possible to
measure the mismatch between the spinor components or its effects
in the transport properties of the charge carriers, this system
would provide an interesting place to probe some predictions of
the Einstein theory of gravitation in two spatial dimensions.

In summary, we have considered the possibility of deforming a
graphene sheet in a conical surface in order to check some
predictions of general relativity in $2+1$ dimensions.  Such a
system makes possible the experimental study of relativistic
massless quasiparticles with charge $e$ on a two-dimensional
surface of a cone, or equivalently, in the ``gravitational field"
(deficit angle) of a ``pointlike particle" of mass $M$ (cone tip).
This surface is locally flat being described by the Minkowski
metric, like the plane. There is no electromagnetic field (such as
that of an infinitesimally thin solenoid [\onlinecite{Jackiw09}])
and the ``gravitational field" is concentrated inside the
pointlike particle and vanishes identically outside of it. From a
naive point of view it seems that in this configuration the
charged quasiparticles could not be affected. Remarkably this is
not the case: they respond to the ``gravitational field" existing
in a place to which they have no access. Therefore, it is similar
to the usual Aharonov-Bohm effect in electromagnetism, but
simulated as a gravitational field by a deficit angle incorporated
in the material. It must have important influences on the
electronic transport properties in ``conical graphene" when
electromagnetic fields are applied. In principle, the effect
proposed here could be detected by interference experiments in
structured materials and consequently these small systems could
provide important probes for general relativity in dimensions
smaller than four.

The authors thank CNPq, FAPEMIG and CAPES (Brazilian agencies) for
financial support.

\end{document}